# Etude mécanique des articles de contention et de leurs effets sur la jambe humaine. Mechanical investigation of compression stockings and of their effects on the human leg.

Stéphane AVRIL[(1)(*)], Sylvain DRAPIER[(2)], Laura BOUTEN[(1)(2)], serge COUZAN[(3)]

[(1)] Ecole Nationale Supérieure des Mines de Saint Etienne, Centre CIS – Département Biomécanique et Biomatériaux, PECM UMR CNRS 5146, 158 Cours Fauriel, 42023 Saint-Étienne Cedex 2, France
[(2)] Ecole Nationale Supérieure des Mines de Saint Etienne, Centre SMS – Département MPE, LTDS UMR CNRS 5513, 158 Cours Fauriel, 42023 Saint-Étienne Cedex 2, France
[(3)] BVSPORT, 104, rue Bergson, 42000 Saint Etienne, France
[(*)] avril@emse.fr

**Résumé :**

Ce papier est une synthèse de différentes études apportant un regard scientifique sur la contention qui consiste à exercer des pressions sur la jambe humaine au moyen de chaussettes ou de bas afin de faciliter le reflux veineux. Il paraît évident que la répartition des pressions sur la jambe affecte le débit sanguin dans les veines. Néanmoins, la répartition conduisant à un débit veineux optimal est non triviale, et diffère en fonction de l'application : traitement médical ou récupération chez les sportifs. Afin d'améliorer les connaissances à ce sujet, un modèle numérique 2D a été mis en place pour calculer les champs de contraintes dans la jambe en prenant en compte le comportement mécanique réel des articles de contention et des tissus de la jambe. Des méthodes d'identification appropriées, basées soit sur le recalage de modèle et la corrélation d'images numériques (pour les articles de contention), soit sur le recalage de modèle par mise en correspondance locale non linéaire d'images fournies par l'IRM (pour les tissus internes de la jambe), ont été développées pour identifier ces paramètres de comportement.

Descripteurs:

*Bas de contention ; tissus mous biologique ; éléments finis (EF) ; hyperélasticité ; corrélation d'images ; recalage de modèle ; recalage d'images médicales ; IRM*

**Abstract:**

This paper presents a synthesis of different studies willing to bring a scientific insight into leg compression, which is the process of applying external compression forces onto the human leg with stockings or socks, for enhancing the venous flow. It seems obvious that the pressure distribution on the leg affects the blood flow in the veins. However, the pressure distribution that leads to the optimal blood flow is not trivial, and it is different with regard to the application: medical treatment or recovering after an effort in sports. In order to improve the scientific knowledge about this topic, a numerical 2D model was set up for computing stress fields inside the leg, accounting for the actual material properties of both the compression stocking and of the leg biological tissues. Suitable identification methods, based either on model updating from digital image correlation (for the compression stockings), or on image warping and model updating from MRI scans (for the internal leg tissues), were developed for retrieving these material properties.

Keywords:

*Compression stocking ; soft biological tissue ; finite elements (FE) ; hyper-elasticity ; model updating ; image correlation ; medical image warping ; MRI*

## 1. Introduction

La contention ou compression élastique (CE) a pour but d'exercer des pressions sur la jambe humaine. Elle permet ainsi de diminuer le volume total de sang veineux et aussi d'augmenter sa vitesse de retour [1]. Les deux applications principales en sont le domaine médical (traditionnels bas à varices) pour traiter certaines pathologies circulatoires, et le sport où les chaussettes de contention sont utilisées pendant ou après un effort de longue durée pour empêcher le sang de s'accumuler dans les jambes (sensations de jambes lourdes) et ainsi faciliter la récupération.

La CE classique prescrite à l'heure actuelle dans le traitement de l'insuffisance veineuse chronique, modérée, importante et sévère, répond au principe de dégressivité : les pressions sont maximales en cheville puis diminuent en remontant vers le haut de la jambe. On parle de "contention dégressive" (CE-Degr). Cependant, le principe de la dégressivité avec des pressions fortes en cheville et plus faibles au mollet est discutable car le volume maximal de sang veineux est contenu dans le mollet, zone où les veines sont quantitativement les plus importantes [2]. Ainsi, afin d'augmenter l'efficacité de la CE, l'application d'un principe inverse de contention dite "contention progressive" (CE-Prog) avec des pressions moindres en cheville et plus fortes au mollet a été mis au point par la société BVSport®, implantée à Saint-Etienne. Il rencontre un large succès chez les sportifs pour la récupération après l'effort. Concernant les applications médicales, une première étude clinique (étude BOOSTER) a montré une sensation de confort supérieure de la CE-prog par rapport à la CE-degr. Une deuxième étude clinique est en cours. Ces études tendent aussi à montrer que la CE-prog favorise la circulation veineuse chez le malade de manière sensiblement meilleure que la CE-degr [3]. De plus, plusieurs études médicales reposant sur l'utilisation de l'IRM ou de l'échographie ultrasonore ont permis de confirmer l'influence significative de la compression du mollet sur le débit veineux [4–6].

Toutefois, aucun modèle mathématique n'a permis à ce jour de comprendre comment les pressions appliquées sur le mollet facilitent le retour veineux. Les études de ce côté-là sont même inexistantes. La CE-degr se justifie par la loi de Poiseuille qui énonce que le débit pour un écoulement laminaire dan une conduite cylindrique est proportionnel au gradient de pression : $Q = \pi R4/8\eta \Delta P/L$, où Q est le débit, R le rayon de la conduite, $\eta$ la viscosité dynamique du fluide, $\Delta P$ le différentiel de pression et L la longueur de la conduite. Ainsi, théoriquement, en appliquant sur la jambe des pressions variant de façon dégressive vers le haut, on augmente le débit en créant un différentiel de pression $\Delta P$. Dans la pratique, des modèles plus complexes prenant en compte les interactions fluide/structure seraient nécessaires car, même si en première approximation l'écoulement du sang dans les veines se fait au repos en régime stationnaire sans déformation dynamique de la paroi des vaisseaux, cela n'est plus valable lorsque le corps est en mouvement. Dans ce dernier cas, les déformations élastiques des parois veineuses et les actions musculaires engendrent localement des accélérations et décélérations du sang. Mais avant d'envisager une étude dynamique, la présente étude va se concentrer sur le cas statique sans action musculaire.

Supposons donc dans cette étude que le débit veineux est bien régulé par la loi de Poiseuille. Il est évident dans ce cas que ce sont les pressions subies directement sur les parois des veines qui vont affecter le débit sanguin, et pas les pressions externes appliquées sur la peau. La relation entre la distribution des pressions externes appliquées sur la peau de la jambe et les pressions internes ressenties sur les parois des veines n'est pas linéaire. De plus, dans la loi de Poiseuille, $\Delta P$ n'est pas le seul paramètre affecté par la compression de la jambe. L'élasticité des tissus engendre une diminution du rayon des veines (paramètre R) sous l'effet de la pression, donc une éventuelle baisse du débit.

Dans cette étude on cherche à déterminer la répartition des pressions hydrostatiques réellement subies dans les tissus constitutifs de la jambe afin de mieux comprendre l'intérêt de la CE-prog par rapport à la CE-degr. Pour cela, un modèle numérique de la jambe humaine sous contention a été développé. Les modèles numériques en biomécanique sont très répandus pour des calculs de structures sur une partie ou la totalité du squelette. Les modèles numériques prenant en compte les tissus mous au niveau de la jambe sont plus rares. On peut citer toutefois à titre d'exemples des modèles du talon [7], de la cuisse chez l'amputé [8, 9], des fesses [10] ou du pénis [11]. Les difficultés inhérentes à la mise en œuvre de tous ces modèles est de déterminer les conditions aux limites et d'alimenter le modèle avec des lois de comportement pertinentes. Concernant le deuxième point, on constate une très grande variabilité des propriétés généralement utilisées dans les modèles, ce qui nécessite de développer pour chaque nouveau modèle une méthode de caractérisation in vivo appropriée. Dans ce papier, une approche permettant de répondre à cette problématique d'identification est proposée.

Le modèle éléments finis (EF) de la jambe sous contention, développé sur le logiciel Zébulon, prend en compte les propriétés mécaniques réelles du textile utilisé pour l'article de contention et les propriétés mécaniques réelles des tissus biologiques de la jambe considérée. Les propriétés mécaniques réelles du textile sont identifiées par recalage à partir de mesures par corrélation d'images des champs de déplacements pour un chargement bi-axial. Concernant les propriétés réelles des tissus biologiques, une méthode inverse originale utilisant l'imagerie médicale par résonance magnétique (IRM) pour recaler un modèle EF et ainsi identifier la loi de comportement a été mise en oeuvre. Après une présentation du modèle et des méthodes d'identification, on montrera les résultats fournis par ce modèle sur un patient donné, puis les perspectives de ces études.

## 2. Présentation du modèle numérique de la jambe sous contention

Un modèle relativement simple de la jambe sous contention a été développé dans le but de valider la démarche globale. La géométrie des différents tissus internes constituant la jambe a été obtenue `a partir d'une image IRM au repos (Fig. 1a). Elle a été segmentée de manière `a définir le contour des différentes zones qui constituent la jambe (voir Fig. 2, avec graisse sur le contour extérieur, muscles au milieu et deux os : tibia et péroné). L'ensemble a été maillé avec 400 éléments triangles linéaires. Le calcul est réalisé dans l'hypothèse des déformations planes. Une étude de convergence a été menée. Les conditions aux limites sont de deux types :

- ❖ en déplacement (Dirichlet), avec un blocage de tous les degrés de liberté sur les contours de deux os (tibia et péroné), ces zones pouvant être considérées comme infiniment rigides par rapport aux tissus mous qui les entourent ;

- ❖ en effort (Neumann) sur le contour extérieur de la jambe. Une pression normale est appliquée sur chaque noeud du maillage. Localement, la valeur de cette pression, notée $p$, est calculée en fonction de la tension locale du textile de contention, notée $T$, selon la loi de Laplace : $p = T/r$, où $r$ est le rayon de courbure local dans la configuration déformée. Les frottements étant négligés, aucune pression tangentielle n'est appliquée. La valeur de $T$ est supposée constante sur tout le contour (frottements négligés). Connaissant l'élongation du textile grâce à la mesure de sa circonférence à la fois au repos et sur la jambe, la valeur de la tension $T$ est directement déduite avec la loi de comportement identifiée sur le textile dans les essais de caractérisation (présentés ci-dessous).

Le comportement des tissus mous de la jambe est supposé hyper-élastique compressible. La compressibilité se justifie par observation de la diminution de volume sur les coupes IRM après application de la contention, comme on le voit sur la figure 2a. Etant donné que les élongations restent faibles, le comportement est simplement modélisé par un modèle Néo-Hookéen [12]. Le potentiel hyper- élastique utilisé est donc :

$$W = C_{10}(I_1 - 3) + K_v[(J_2 - 1)/2 - \ln(J)]$$

où : $J = \det(F)$ et $I_1 = J^{-2/3}\mathrm{tr}(^tF.F)$ avec $F$ désignant le gradient de la transformation et tr la trace.

Des propriétés différentes sont considérées pour la graisse (parties claires sur les IRM de la Fig. 1) et pour les muscles (parties sombres sur les IRM de la Fig. 1). Le comportement de chaque tissu mou est caractérisé par les deux constantes : $C_{10}$ et $K_v$. Ce premier modèle numérique très simpliste de la jambe sous contention est donc piloté par quatre paramètres de comportement. Il existe une très grande variabilité des valeurs de ces paramètres fournies dans la littérature [8–11]. De manière à utiliser des valeurs pertinentes en fonction de chaque patient étudié, une procédure d'identification in vivo de ces paramètres a été développée. Elle repose sur le recalage du modèle numérique grâce aux coupes faites en IRM. Cette procédure de recalage est décrite à la partie 4. Lorsque les quatre paramètres de comportement sont identifiés, la résolution de l'équilibre mécanique se fait de manière incrémentale sur 10 itérations. L'objectif final du calcul sur la jambe consiste à calculer le champ de pression hydrostatique sur toute la coupe de la jambe (exemple de résultat donné sur Fig. 5).

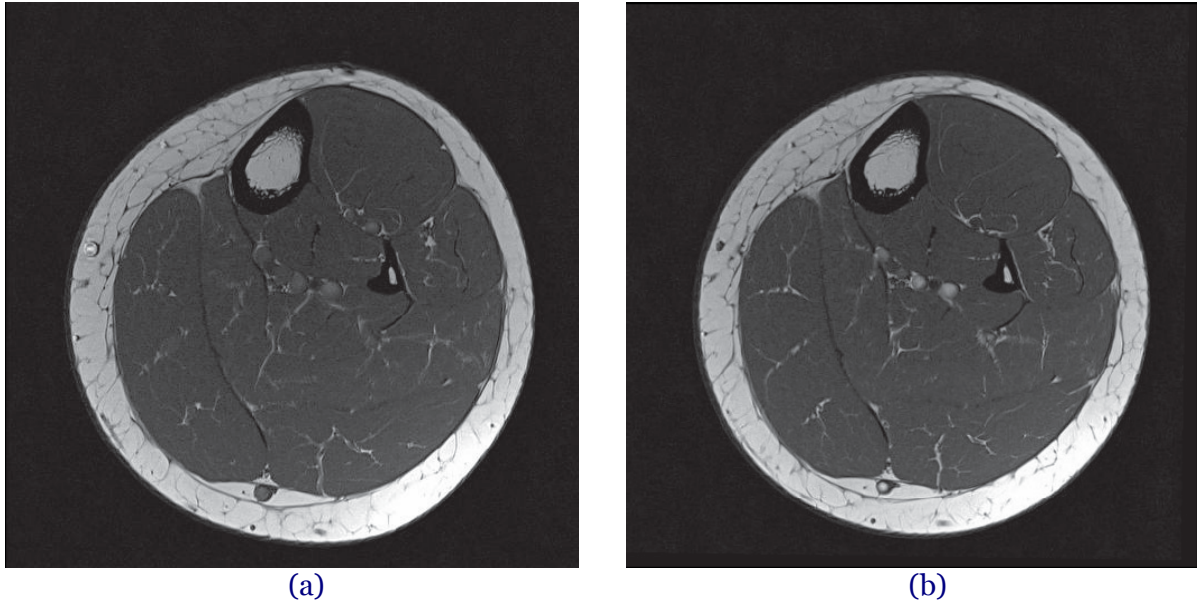

*Figure 1 : Coupes de la jambe réalisée par IRM. (a) Avant port de la chaussette de contention. (b) Pendant le port de la chaussette de contention.*

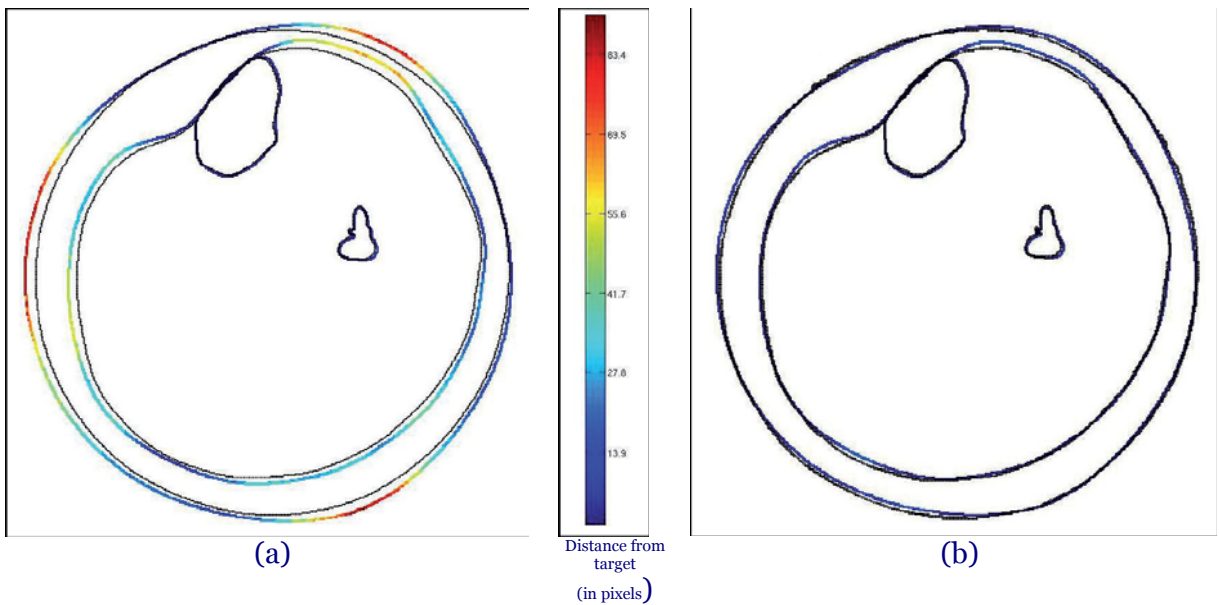

*Figure 2 : Superposition des contours cibles et des contours à recaler. (a) Avant transformation par le modèle numérique. (b) Après transformation par le modèle numérique.*

## 3. Identification du comportement bi-axial des articles de contention

L'identification de la réponse des textiles (tricots) constituant les articles de contention est nécessaire pour fournir les conditions aux limites de Neumann du modèle de la jambe sous contention. L'identification de la réponse de ce type de matériau fibreux souple travaillant en grandes déformations est une problématique récente et complexe. Les difficultés qui se posent sont multiples, elles proviennent de la structure même du tricot, milieu discontinu par nature, mais aussi des faibles niveaux d'efforts engendrés, autant que du type de fil élastomérique utilisé qui génère une réponse viscoélastique ou encore du domaine d'utilisation en grandes déformations, entre 60 % et 120 % de déformation dans le sens trame i.e. dans le sens circonférentiel de la jambe. Ainsi, pour répondre à ces difficultés, une méthodologie d'identification originale du comportement bi-axial en grandes déformations des textiles élastomériques a été développée. Elle a d'abord été appliquée sur des articles de CE-degr dans le cadre d'un travail antérieur à la présente étude [13–15]. Cette méthodologie s'appuie en premier lieu sur un montage spécifique [15] permettant l'acquisition d'images successives représentant l'état du textile au cours de sa sollicitation. Pour ces essais, des cyclages sont réalisés avant les mesures, conformément à la pré-norme [16]. Les séquences d'images acquises sont ensuite traitées par corrélation d'images numériques [17], la granularité du textile servant de mouchetis naturel, afin d'en déduire des champs de déplacements. La sollicitation bi-axiale est ici induite par une pré-tension imposée dans le sens transverse à la sollicitation. Sous l'effet de cette pré-tension, le fort coefficient de liage induit une contraction transverse bloquée, ce qui induit la réponse bi-axiale souhaitée. En parallèle, un modèle numérique EF du tricot a été mis en place, dans le cadre d'une formulation corrotationnelle, à l'aide du code de calcul Zebulon, permettant de simuler la réponse du tricot sous le chargement précédemment décrit. Cette réponse simulée est recalée par rapport aux déplacements mesurés, permettant ainsi d'identifier la réponse bi-axiale du textile en grandes déformations tel que présenté sur la figure 4. On note que pour les déformations utiles (déformation sens trame [60 %, 100 %] et déformation sens chaîne [20 %, 40 %]), la surface hypo-élastique polynômiale représente parfaitement la réponse expérimentale. Pour les déformations extrêmes, un écart subsiste, mais ces déformations sont hors des déformations utiles.

La courbe de réponse obtenue nous permet d'alimenter les valeurs de T dans la loi de Laplace qui est utilisée pour définir les contions aux limites en pression du modèle de la jambe sous contention. Seule la réponse uni-axiale dans le sens trame est nécessaire pour le modèle 2D de la jambe sous contention présenté dans ce papier. Toutefois, la caractérisation de la réponse bi-axiale a été faite ici dans la perspective de développer, à moyen terme, un modèle 3D de la jambe sous contention. Une étude complémentaire a également été menée afin de vérifier la pertinence de la loi de Laplace pour relier la tension locale du textile à la pression externe appliquée sur la peau de la jambe [13, 15]. Un modèle physique de jambe rigide a permis de recouper les mesures réalisées expérimentalement à l'aide de capteurs de pression spécifiques et les résultats de pression calculés par des simulations EF (Abaqus$^{®}$) en 2D simulant le contact du textile sur la jambe. Dans ces simulations, la loi de comportement uni-axial implémentée dans une subroutine UMAT a été identifiée à partir d'essais uni-axiaux quasi-statiques, après avoir vérifié que le temps de relaxation du tricot était environ 5 fois plus élevé que la durée des tests. Les résultats des simulations, ainsi que les mesures, montrent que la pression locale est bien liée à la courbure locale de la jambe ('*Laplace locale*' sur Fig. 3) et valident donc l'utilisation de la loi de Laplace pour définir les conditions aux limites du modèle de la jambe sous contention.

## 4. Identification de la loi de comportement des tissus mous biologiques de la jambe

On rencontre fréquemment dans la littérature des comparaisons entre les champs de déplacements numériques fournis par un calcul EF et les champs expérimentaux fournis par une méthode comme la corrélation d'images [17]. On peut même recaler le modèle EF en prenant certains paramètres de comportement comme variables de contrôle, puis en minimisant l'écart entre les champs expérimentaux et numériques, comme cela a été fait pour déterminer la loi de comportement du textile de contention. Ce type d'approche pourrait être utilisé ici pour identifier la loi de comportement des tissus biologiques de la jambe si on était capable de mesurer expérimentalement les déplacements à l'intérieur de la jambe entre l'état libre et l'état sous contention. Malheureusement, bien que nous disposions d'images de la jambe avant et après application de la contention, la corrélation d'images classiques (corrélation par sous-imagettes [17]) ne permet pas d'en déduire des champs de déplacement suffisamment précis car les images dont nous disposons ne sont pas d'un contraste suffisant (pas de mouchetis naturel). Une tentative d'utilisation de la corrélation par sous imagettes sur des coupes IRM de la jambe est présentée dans [18], montrant une forte variabilité des valeurs de déplacement localement obtenues.

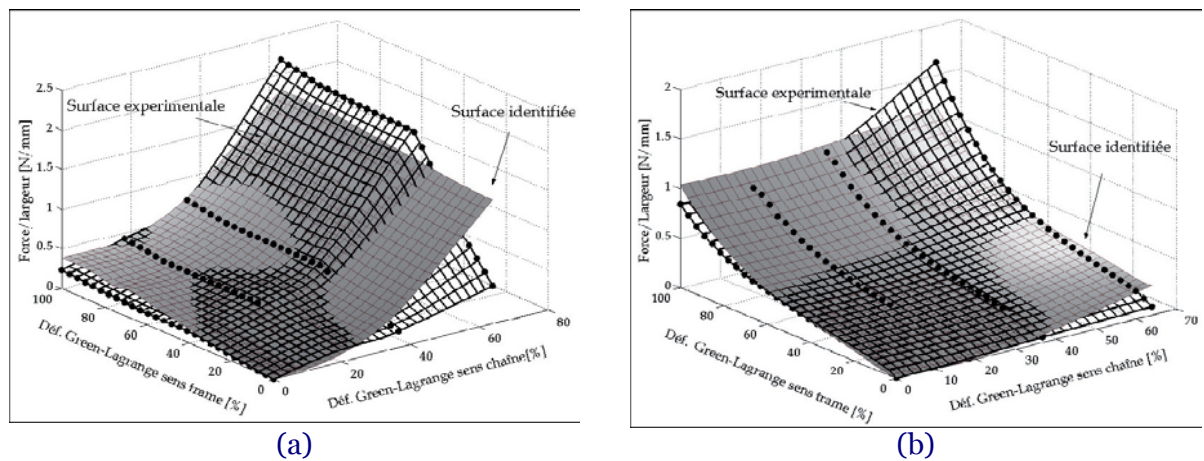

(a) (b)

*Figure 4 : Surface de réponse (a) sens trame et (b) sens chaîne d'un tricot (référence industrielle) : mesures expérimentales et réponse identifiée [14, 15].*

L'idée est donc de procéder à de la corrélation d'image *globale*, qui est encore faiblement utilisée dans la littérature [19]. Le principe est le suivant : au lieu de déterminer localement les translations qui permettent de passer de l'image initiale (IRM sans port de la chaussette) à l'image finale (IRM avec port de la chaussette), on définit de manière paramétrée une transformation globale permettant de passer d'une image à l'autre [20] et on détermine les paramètres de cette transformation par minimisation au sens des moindres carrés sur l'écart entre les intensités d'images.

Dans notre cas, la transformation globale est celle calculée par le modèle EF de la jambe sous contention. Les paramètres de cette transformation sont donc les 4 propriétés mécaniques ($C_{10}$ et $K_v$ pour les muscles et la graisse). On recale le modèle en utilisant la transformation géométrique calculée pour déformer l'image initiale et en cherchant à la faire s'apparier au mieux avec l'image finale. Au lieu d'utiliser tous les voxels comme dans [20], seuls les contours des muscles et de la graisse sont utilisés ici. Ces contours principaux sont déterminés par segmentation des coupes IRM. Les contours cibles (en noir sur la chaussette (état chargé). Le contour à recaler est celui de l'état non chargé (contour en couleur sur Fig. 2a). Pour chaque calcul EF du modèle de la jambe sous contention, les déplacements interpolés aux pixels concernés sont appliqués pour déformer le contour à recaler. Un exemple de contour déformé est donné sur figure 2b (contours en couleur). L'échelle de couleur correspond à la distance locale entre le contour déformé et le contour cible. Le recalage se fait par minimisation, la fonction coût étant définie comme la distance moyenne entre le contour déformé et le contour cible. La minimisation de cette fonction coût est faite par une méthode de simplexe. La convergence de cette méthode a d'abord été validée sur un cas test numérique. Elle a ensuite été appliquée pour identifier les paramètres du modèle des tissus de la jambe sur un patient donné.

## 5. Résultats et discussion

Les contours en couleur de la Figure 2b correspondent aux contours déformés pour le minimum de la fonction coût. Les paramètres identifiés sont : $C_{10}$ = 8,5 kPa et $K_v$ = 42,9 kPa pour la graisse et $C_{10}$ = 12,9 kPa et $K_v$ = 43,8 kPa pour les muscles. Un appariement parfait sur tout le long du contour n'a pas pu être obtenu. Les raisons principales proviennent de certaines hypothèses trop fortes liées au modèle, en particulier vis-à-vis de la partition des tissus mous de la jambe en deux matériaux différents seulement : graisse et muscle. Les hétérogénéités locales de la jambe ne sont pas prises en compte, par exemple au niveau de la veine grande saphène, et c'est à cet endroit que sont observés les écarts les plus forts entre contours cibles et recalés.

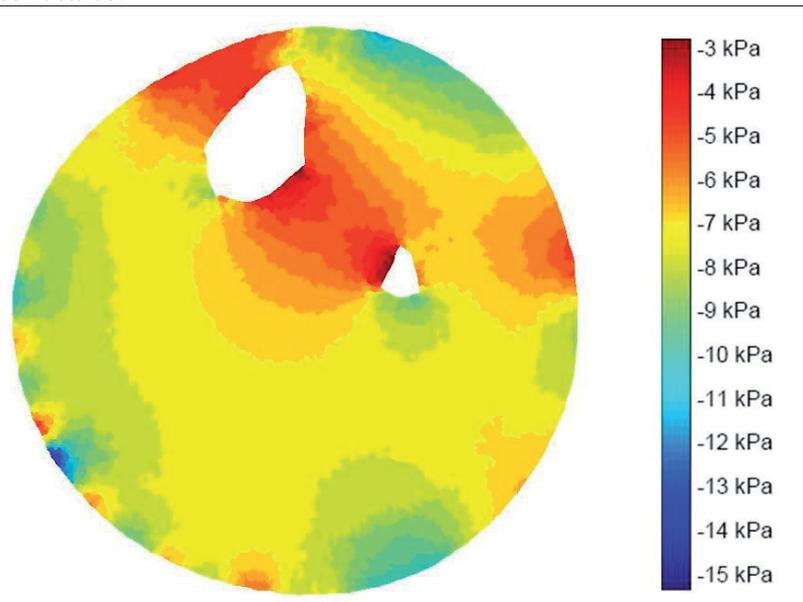

*Figure 5 : Pressions hydrostatiques calculées dans chaque tissu de la jambe après recalage du modèle pour un patient et une chaussette de contention donnés.*

Une fois le modèle recalé et les propriétés mécaniques des tissus mous identifiées, le champ de pression au sein des tissus de la jambe peut être obtenu en réalisant un calcul avec les paramètres identifiés (Fig. 5). Ces résultats sont parmi les premiers obtenus et leur interprétation est toujours en cours. On constate une hétérogénéité importante des pressions. Les tissus en profondeur subissent les pressions les moins élevées. Les hétérogénéités proches de la surface sont dues à l'existence des variations de courbures, induisant des pressions locales plus élevées d'après la loi de Laplace.

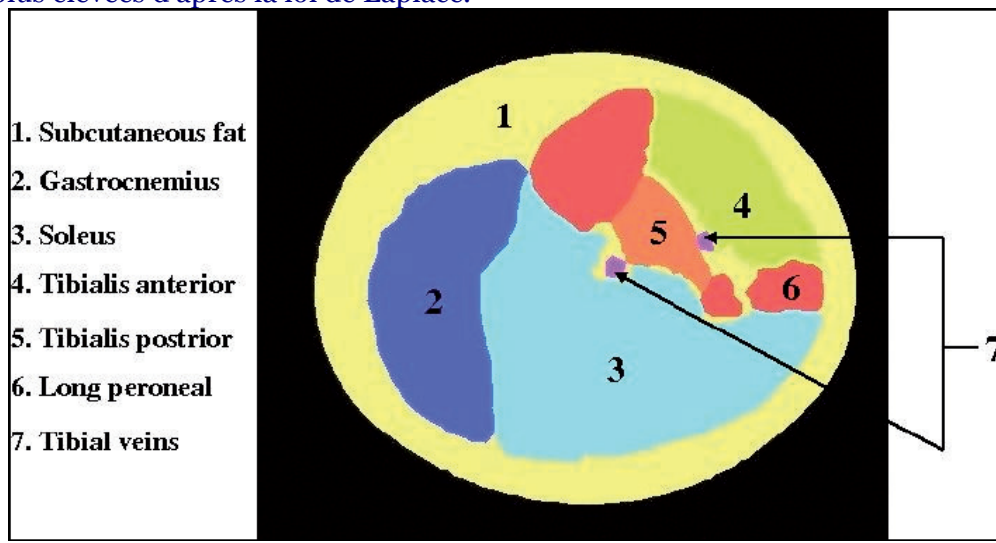

*Figure 6 : Principaux tissus mous de la jambe humaine.*

Les champs de pression fournis par le modèle sont prometteurs en termes d'applications. Il sera intéressant d'analyser les différences entre différentes stations (debout, couché). De plus, les résultats fournis par ce modèle vont pouvoir être corrélés avec les résultats fournis par l'appareil de mesure de pression veineuse AMPVE® [6]. Cela permettra de déterminer quelle est la pression locale permettant de réduire le diamètre des principales veines de la jambe.

Cela permet aussi de déterminer quels sont les muscles qui subissent le moins de compression et donc au niveau desquels l'effet de facilitation du retour veineux sera le moins efficace.

## 6. Conclusions

Afin de mettre en oeuvre un premier modèle mécanique simpliste permettant de déterminer les pressions subies à l'intérieur de la jambe humaine sous compression élastique, des méthodes d'identification pour matériaux souples et fibreux ont été utilisées. Basées sur l'imagerie, elles ont permis un recalage du modèle. Les premières applications du modèle sont en cours, concernant principalement un recoupement avec des résultats médicaux déjà observés, en particulier concernant l'accumulation de sang plus importante dans certaines parties du mollet plutôt qu'ailleurs.

Le modèle de la jambe sous contention présenté dans cet article est très simpliste étant donné l'objectif premier qui était de valider la démarche générale, y compris la méthode d'identification in vivo des tissus mous biologiques. A court terme, le passage à une modélisation 3D est envisagé, afin d'intégrer la loi de comportement biaxiale du textile et pour ne plus dépendre de l'hypothèse très restrictive des déformations planes dans la jambe. La prise en compte d'un plus grand nombre de constituants dans la jambe pourra alors être envisagée (Fig. 6). A plus long terme, le couplage avec des calculs d'écoulement sanguin dans les veines, avec interactions fluide-structure, sera traité.

## Références